\documentstyle[12pt]{article}
\normalsize

\let\oldtheequation=\theequation
\def\doteqs#1{\setcounter{equation}{0}
            \def\theequation{{#1}.\oldtheequation}}

\newcounter{sxn}
\def\sx#1{\addtocounter{sxn}{1} \bigskip\medskip \goodbreak \noindent{\large\bf
\centerline{\thesxn.~~#1}} \nobreak \medskip}
\def\sxn#1{\sx{#1} \doteqs{\thesxn}}

\newcounter{axn}
\def\ax#1{\addtocounter{axn}{1} \bigskip\medskip \goodbreak \noindent{\large\bf
{\Alph{axn}.~~#1}} \nobreak \medskip}
\def\axn#1{\ax{#1} \doteqs{\Alph{axn}}}

\def\br{\begin{thebibliography}{abc}}
\def\er{\end{thebibliography}}

\date{}

\begin{document}
\bibliographystyle{unsrt}
\footskip 1.0cm
\thispagestyle{empty}
\setcounter{page}{0}
\begin{flushright}
UAHEP-931 \\
Jan 1993
\end{flushright}
\begin{center}{\LARGE LAGRANGIAN AND HAMILTONIAN FORMALISM  \\
  ON A QUANTUM PLANE\\ }
\vspace*{6mm}
{\large  M. Lukin, A. Stern and I. Yakushin\\}
\newcommand{\bc}{\begin{center}}
\newcommand{\ec}{\end{center}}
\vspace*{5mm}
{\it Department of Physics, University of Alabama, \\
Tuscaloosa, AL 35487, USA.}\ec
\bigskip
\normalsize
\centerline{\bf ABSTRACT}
We examine the problem of defining Lagrangian and Hamiltonian mechanics
for a particle moving on a quantum plane $Q_{q,p}$.  For Lagrangian
mechanics, we first define a tangent quantum plane $TQ_{q,p}$ spanned by
noncommuting  particle coordinates and velocities.
Using techniques similar to those of Wess and Zumino, we construct two
 different differential calculi on $TQ_{q,p}$.  These two
differential calculi can in principle give rise to two different particle
dynamics, starting from a single Lagrangian.  For Hamiltonian
mechanics, we define a phase space $T^*Q_{q,p}$ spanned by noncommuting
 particle coordinates and momenta.
The commutation relations for the momenta
can be determined only after knowing their functional
 dependence on coordinates and velocities.
 Thus these commutation relations, as well as the
 differential calculus on $T^*Q_{q,p}$, depend on the initial choice
 of Lagrangian.  We obtain the deformed Hamilton's
 equations of motion and the deformed Poisson brackets, and their
 definitions also depend on our initial choice of Lagrangian.
We illustrate these
ideas for two sample Lagrangians.  The first system we examine
 corresponds to that of a nonrelativistic
particle in a scalar potential.
The other Lagrangian we consider is first order in time derivatives
and it is invariant under the action of the
 quantum group $SL_q(2)$.  For that system, $SL_q(2)$ is shown to
 correspond to a canonical symmetry transformation.
\newpage
\newcommand{\be}{\begin{equation}}
\newcommand{\ee}{\end{equation}}
\newcommand{\ba}{\begin{eqnarray}}
\newcommand{\ea}{\end{eqnarray}}
\newcommand{\no}{\nonumber}
\baselineskip=24pt

\sxn{Introduction}

Recently, quantum groups\cite{rev} have attracted much attention in the
physics literature partially due to their
possible utility in describing particles in two spatial
dimensions with generalized spin and statistics\cite{YG,ow}.
Quantum groups have the property that they act covariantly on generalized
spaces, known as quantum planes\cite{man},
which are parametrized by noncommuting coordinates.
These coordinates have been viewed
as the classical limit of ``deformed" creation and annihilation
operators\cite{bf}.  Such creation and annihilation
operators have, in turn, been utilized in the
construction of quantum group generators\cite{bied,bied2}

Now if quantum plane coordinates are
the classical limit of deformed creation and annihilation
operators, how can ``classical" dynamics be introduced
on a quantum plane, such that upon canonical quantization,
one recovers the deformed creation and annihilation operators?
More generally, we can ask how does one define classical Lagrangian and
Hamiltonian mechanics on a quantum plane?  This is similar to
the question of defining classical dynamics for particles moving
on spaces parametrized by Grassmann variables; a question
 which was answered long ago\cite{cas}.

In this article, we write down the Lagrangian and Hamiltonian
formalism for a particle moving on the
two dimensional quantum plane $Q_{q,p}$\cite{soni}.
$Q_{q,p}$ is a two parameter deformation of the ``plane",
the parameters being $q$ and $p$.
If $x$ and $y$ denote the noncommuting coordinates of $Q_{q,p}$, then
one of the parameters enters in their commutation relation:
\be
xy = q \;yx     \quad.
\ee

The other parameter $p$ appears in the
differential calculus on $Q_{q,p}$.  (The differential calculus
for the one parameter quantum plane was written down
by Wess and Zumino\cite{wz}, and generalized
for multiparameter quantum planes in refs.\cite{soni}.)
For this, one introduces one forms $dx$ and $dy$, where
$d$ denotes an exterior derivative
and, as usual, $d^2=0$.  Unlike the case with commuting
coordinates, the exterior product of two one forms need not be
antisymmetric.  Instead,
\be
dx \wedge  dy =-{1\over p}\; dy \wedge  dx   \;.
 \ee

The commutation relations (1.1) and (1.2) have the feature
that they are unchanged in form under the action of a two parameter
deformation $GL_{q,p}(2)$ of the general linear group in
two dimensions.  Under the action of $GL_{q,p}(2)$,
\be
\pmatrix{x  \cr
y \cr  } \rightarrow
\pmatrix{x'  \cr
y' \cr  }
=[ T ]  \;
\pmatrix{x  \cr
y \cr  }   \quad ,
\ee
where $[ T ]$ denotes a $2\times 2$ matrix
 $[ T ]=  \pmatrix{ A & B \cr  C & D \cr}$,
whose the matrix elements $A,B,C,$ and $D$ commute with the
coordinates $x$ and $y$, but do not commute
amongst themselves.  Rather, they satisfy
$$
AB=p\;BA\;, \qquad AC=q\;CA\;,
$$
\be
CD=p\;DC\;,\qquad  BD=q\;DB\;,
\ee
$$
BC={q\over p}\;CB\;,\quad
AD-DA=(q-{1\over p})\;CB\;.
$$

To completely define the differential calculus on the quantum plane
it is necessary to specify the commutation relations between
coordinates $x$ and $y$ and their exterior derivatives  $dx$ and $dy$
 in a consistent manner.  According to Ref.\cite{soni}, one can choose
\ba
x\;dx& =&pq\; dx\; x \;,     \no\\
x\;dy&=& q\;dy\; x + (pq-1)\; dx\; y \;,    \no\\
y\;dx&=&p\; dx\; y \;,\\
y\;dy&=&pq\; dy\; y \;. \no
\ea
The relations (1.1), (1.2) and (1.5) are consistent in that i)
 it follows that from (1.5) that $d(xy - q \;yx)=0 $
( we assume the usual Leibniz rule for the exterior derivative),
ii) the exterior derivative of the equations (1.5)
agrees with eq. (1.2), and iii) no secondary conditions arise from
commuting $dx$ or $dy$ through the left-hand side of $xy - q \;yx=0 $.
Provided $pq\ne -1$, one further finds that  $(dx)^2 =(dy)^2=0$.
iv) It can also be checked that like (1.1) and (1.2),
the relations (1.5) are preserved under the
$GL_{q,p}(2)$ transformations (1.3).
(The consistency conditions can be reexpressed in terms of
Yang-Baxter equations\cite{wz}.)  v) Finally we note that
the standard differential calculus on a plane is recovered
in the limit $q=p=1$.

[Alternative commutation relations can be found which satisfy i-v).
They are obtained by making the replacement
$x\rightarrow y,\; y\rightarrow x,\;   q \rightarrow q^{-1},\;
p \rightarrow p^{-1}\;$ in the relations (1.5).  We will not consider the
alternative solutions here, but expect that they lead to
conclusions which are similar to those we obtain starting from (1.5).]

Now to specify the motion of a particle on $Q_{q,p}$, we need to
introduce a velocity vector $v=(\dot x,\dot y)$, and define
the commutativity properties of $\dot x$ and $\dot y$.  We shall
do this in Section 2.  $x,\;y,\;\dot x$ and $\dot y$
coordinatize the ``tangent quantum plane" which we shall denote by
 $TQ_{q,p}$.  The Lagrangian will be a function of these variables.
In order to write down the Lagrangian formalism, we must be able
to perform variations on $TQ_{q,p}$, and consequently
define a differential calculus on $TQ_{q,p}$.
Thus not only do we need to determine the commutativity properties
$x,\;y,\;\dot x$ and $\dot y$, we also need to determine the
commutativity properties of their variations.
This will be done in Section 3 and in the Appendix.  We shall require,
as usual, that the time derivative commutes with the variational
derivative.  (For us, time will be a commuting parameter.)
In addition, we require that the differential calculus
is preserved under the $GL_{q,p}(2)$ transformations (1.3).  Two
distinct solutions are shown to be consistent with these
requirements.  In principle, they can lead to two distinct particle
dynamics - starting from a single Lagrangian [and from a differential
calculus on $Q_{q,p}$ defined by (1.1), (1.2) and (1.5)].

The Lagrangian and Hamiltonian formalisms are developed in Section 4.
For the former, we get the usual form for the Euler-Lagrange
equations of motion.  For the latter, we need to define a phase
space $T^*Q_{q,p}$ spanned by the particle coordinates and momenta.
Now the commutation relations for the momenta
can only be determined after knowing their functional
 dependence on coordinates and velocities.
 Therefore these commutation relations depend on the initial choice
 of Lagrangian.  For the Hamiltonian formalism, we must be able
to perform variations on $T^*Q_{q,p}$, and thus
define a differential calculus on $T^*Q_{q,p}$.
The commutation relations associated with this differential calculus will
also depend on the initial choice of Lagrangian.  We further find that
 the form of the ``deformed" Hamilton's equations of motion and
 the ``deformed Poisson brackets" will depend on this choice, as well.
   The deformed Poisson brackets coincide with
 the usual Poisson brackets for commuting variables in the limit $q=p=1$.
 Similar deformed Poisson brackets have been postulated by
several authors\cite{pb}, and they are the classical analogue
of deformed commutators or ``quommutators" appearing in quantum theory.
To define the deformed Poisson bracket, we require
that the time derivative of any function on $T^*Q_{q,p}$ is the deformed
Poisson bracket of that function with the Hamiltonian.

We illustrate the above described formalism by studying for two sample
Lagrangians.  The first Lagrangian we examine is quadratic in time
derivatives and, in the limit $q=p=1$, describes a nonrelativistic
particle in a scalar potential.  The commutativity properties of
the coordinates and velocities restricts possible choices
for the potential.  The second Lagrangian we examine is first
order in time derivatives.  It therefore has constraints in the
Hamiltonian formalism, and they can be eliminated using the analogue of
the Dirac prescription.  The Lagrangian is invariant under the
action of the quantum group $SL_q(2)$, and these
transformations are shown to preserve the Dirac brackets.
They therefore correspond to canonical transformations.

We remark in Section 5, that by quantizing the second Lagrangian
one may indeed obtain deformed creation and annihilation operators.
But this result is not general as it may not apply for other Lagrangian
systems.   Additional concluding remarks are made in Section 5.

\sxn{The Tangent Quantum Plane $TQ_{q,p}$}

We wish to describe the dynamics of a particle moving
on the quantum plane.  For this, let us
 parametrize the particle trajectory by a (commuting) time
coordinate $t$.  Let us also
introduce a velocity vector $v=(\dot x,\dot y)$ for the particle,
the dot denoting a time derivative.
$x,\;y,\;\dot x$ and $\dot y$
coordinatize the ``tangent quantum plane" $TQ_{q,p}$.
To fully define $TQ_{q,p}$, we must specify the commutation
relations for the velocity components
with the coordinates $x$ and $y$, as well as, with themselves.
For this, we write the one forms $dx$ and $dy$
 on $TQ_{q,p}$ in terms of the velocity components
$ dx=\dot x\; dt   \quad{\rm and}\quad
dy=\dot y \; dt \;.    $
Since we are defining $t$ to be a commuting parameter,
the commutation relations of
$x$ and $y$ with velocity components $\dot x$ and $\dot y$ must be
 the same as
the commutation relations of $x$ and $y$ with $d x$ and $d y$,
respectively.  That is
\ba
x\;\dot x& =&pq\; \dot x\; x \;, \no\\
x\;\dot y&=&  q\;\dot y\; x   +(pq-1)\; \dot x\; y\;, \no\\
y\;\dot x&=&p\; \dot x\; y \;,\\
y\;\dot y&=&pq\; \dot y\; y \;. \no
\ea
These relations are preserved under $GL_{q,p}(2)$ transformations.

It remains to specify the commutation relation between
$\dot x$ and $\dot y$.  We can start with the general ansatz:
$
\dot x \;\dot y = c_1\; y \; x  +c_2\; \dot x \; x +c_3 \;\dot y \; y
+ c_4 \;\dot x \; y  +c_5\; \dot y \;x +c_6\;  \dot y\;\dot x     \;.
$
{}From the requirement that no secondary conditions result from commuting
$\dot x$ and $\dot y$ through eqs. (2.1), we must have
$c_1=c_2=c_3=c_4=c_5=0$ and $c_6=q$.  We are thus left with
\be
\dot x \;\dot y =  q\;  \dot y \dot x     \quad.
\ee
This relation too is preserved under transformations (1.3).

The commutation relations for the coordinates
$\{z_1=x,\;z_2=y,\;z_3=\dot x,\;z_4=\dot y\;\}$
of the four dimensional quantum manifold $TQ_{q,p}$
are specified by eqs. (1.1), (2.1) and (2.2).  We note that
these relations ${\it cannot}$ be written in the simple form:
$z_i z_j= q_{ij}z_j z_i\;,\;\; i<j$.

By taking the time derivatives of the relations (2.1) and (2.2),
one obtains the commutativity properties for higher order derivatives
of $x$ and $y$.  For example, the derivative of relations (2.1) leads to
the commutation relations for components of acceleration with the
coordinates:
\ba
x\;\ddot x& =&pq\; \ddot x\; x +(pq-1)\;{\dot x}^2\;, \no\\
x\;\ddot y&=&  q\;\ddot y\; x
  +(pq-1)(q\; \dot y \;\dot x\;+\;\ddot x \;y)\;, \no\\
y\;\ddot x&=&p\; \ddot x\; y
   +(pq-1)\; \dot y\; \dot x\;\;,\\
y\;\ddot y&=&pq\; \ddot y\; y
 +(pq-1)\;{\dot y}^2 \;. \no
\ea

For $pq\ne 1$, these conditions alone put severe restrictions
on the allowable trajectories on $Q_{q,p}$.
For example, for a ``free particle"
$\ddot x =\ddot y =0$, eqs. (2.3) imply that the velocity
components are nilpotent
 ${\dot x}^2 ={\dot y}^2 =\dot x \dot y = 0$.
Similar conclusions follow for a particle in a potential
 as we illustrate in Section 4.

To obtain nontrivial particle dynamics it may be desirable to impose
that $pq=1$.  In that case we get no new restrictions on a free
particle.   For the sake of generality, we shall not set
 $pq=1$ in the discussions which follow, except where otherwise stated.

\newpage
\sxn{Differential Calculus on $TQ_{q,p}$}

For Lagrangian mechanics we must be able
to perform variations on $TQ_{q,p}$, and thus we must
define a differential
calculus on $TQ_{q,p}$.  That is, we need to know the commutation
relations between coordinates $z_i$ of $TQ_{q,p}$
and their exterior (or variational)
derivatives, which we now denote by $\delta z_i$.
(We define the exterior derivative $\delta$ to be equivalent to
$ d$, when it acts on $Q_{q,p} \subset  TQ_{q,p}$.)
In what follows we find that it is possible to define
two different kinds of differential calculi on $TQ_{q,p}$.

As is usual in Lagrangian mechanics, we shall assume that
the time derivative commutes with the variational derivative.
A few commutation relations necessarily follow from this.
They are obtained by comparing the
 time derivative of (1.5) with the exterior derivative
of (2.1).  This leads to:
\ba
 \dot x\; \delta x &=&\delta x\;\dot x  \;,    \no\\
 \dot y\; \delta y &=&\delta y\;\dot y  \;,    \no\\
  x\; \delta \dot x &=&pq\;
  \delta \dot x\; x+(pq-1)\;\delta x\; \dot x  \;,   \\
  y\; \delta \dot y &=&pq\;
  \delta \dot y\; y+(pq-1)\;\delta y\; \dot y  \;,    \no
\ea
where we assume the Leibniz rule for $\delta$ and $pq\ne -1$.
In addition to (3.1), one obtains the identities:
\ba
 \dot y\; \delta x +y\;\delta \dot x
  &=&p(\delta \dot x\; y +\delta x \;\dot y) \;,    \no\\
\dot y\; \delta x + p\;\dot x\;\delta y& =&
p \;\delta x \;\dot y +\delta y \;\dot x\;, \\
\dot x \;\delta y + x\; \delta\dot y&=&q(\delta\dot y\;
 x + \delta y \;\dot x)
+(qp-1)(\delta \dot x\; y +\delta x \;\dot y)\;.\no
\ea

By further taking the exterior derivative of (3.1) and (3.2),
we get commutation relations between one forms
 $\delta z_i$ on  $TQ_{q,p}$:
 \ba
 \delta x \wedge  \delta \dot x &=&-\delta \dot x \wedge  \delta x   \;,
 \no\\
\delta y \wedge  \delta \dot y &=&-\delta \dot y \wedge  \delta y    \;,
\ea
along with the identity:
\be
\delta  x \wedge  \delta \dot y
+{1\over p}\delta  y \wedge  \delta\dot x
=-\delta \dot x \wedge  \delta y -
  {1\over p}\; \delta \dot y \wedge  \delta x \;.
 \ee
As in the previous section,
we have assumed that $\delta^2=0$ and $(\delta z_i)^2=0$.

To completely define the differential calculus on
 $TQ_{q,p}$, we shall make ans\"atze for the remaining
 commutation relations.  Our ans\"atze are such that all terms involve
the same number of velocities and coordinates of $Q_{q,p}$.

We start with:
$
\delta\dot x\wedge\delta \dot y =r\;\delta\dot y\wedge\delta \dot x\;,
$
$r$ being a c-number.
Invariance under $GL_{q,p}(2)$ transformations (1.3) immediately
fixes $r$ to be either $r=q$
or $r=-{1\over p}$.  With the former solution, however, we do
 not get the usual antisymmetric exterior product
in the limit $q=p=1$.  We shall therefore not consider this case.
We are thus left with
\be
\delta\dot x\wedge\delta \dot y =
-{1\over p}\;\delta\dot y\wedge\delta \dot x\;.
\ee

Eqs. (1.2), (3.3) and (3.5) give some of
the commutation relations between two one forms $\delta z_i$ on
$TQ_{q,p}$.  For the remaining such relations, we define a
 $2\times 2$ matrix
$[ f  ]  =\pmatrix{ f_{11} & f_{12}\cr f_{21} & f_{22}\cr}$, with
the elements $f_{ij}$ being c-numbers, and
\be
 \pmatrix{\delta x\wedge\delta \dot y \cr
\delta y\wedge\delta \dot x \cr   }
= [ f ]  \;
 \pmatrix{\delta \dot y \wedge\delta x\cr
\delta \dot x \wedge\delta y\cr  }    \quad .
\ee
The four matrix elements of $[ f ] $ can be determined I) by demanding
that no secondary conditions on combinations of
 $\delta z_i$ result from
commuting $\delta z_i$ through the relation (1.2),
II) from the identity (3.4) and III) by demanding that the relations
(3.6) are preserved under the $GL_{q,p}(2)$ transformations (1.3).
(The consistency requirements can be formulated in terms of
Yang-Baxter equations\cite{wz}.)

I) By multiplying $\delta \dot x$  on the right of
$\delta x \wedge  \delta y +{1\over p}\; \delta y \wedge  \delta x
=0$, we find
\be
f_{12}f_{21}=0\;.
\ee

II)  Upon substituting the ansatz (3.6) into the identity (3.4),
 we get the conditions:
\be
pf_{11}+f_{21}=-1\quad{\rm and}\quad
pf_{12}+f_{22}=-p\; \;.
\ee

III) Invariance under $GL_{q,p}(2)$ transformations is insured provided
$$
f_{11}+qf_{12}=-q\;,\quad
pf_{11}-qf_{22}=0
$$
\be
{\rm and}\quad (qp-1)f_{11}-qf_{12}+qf_{21}=0\;.
\ee
For general $q$ and $p$, there are two possible
solutions to eqs. (3.7-9), which we shall call
case a) and case b).  The case a) commutation relations are:
\ba
 \delta x\wedge\delta \dot y &=&-{1\over p}\;\delta\dot y\wedge\delta x
+ {1\over{pq}}(1-pq)\;\delta \dot x \wedge\delta y \;,\no\\
\delta y\wedge\delta \dot x &=&-{1\over q}\;\delta \dot x \wedge\delta y
\;,
\ea
while for case b) we have:
\ba
 \delta x\wedge\delta \dot y &=&-q\;\delta\dot y\wedge\delta x \;,\no\\
\delta y\wedge\delta \dot x &=&(pq-1)\;\delta\dot y\wedge\delta x
-p\;\delta \dot x \wedge\delta y  \;.
\ea

Lastly, eight commutation relations remain to be specified in order to
completely fix the calculus.  They are the
commutation relations between coordinates $z_i$ of $TQ_{q,p}$
and one forms $\delta z_i$ on $TQ_{q,p}$ not already contained in eqs.
(1.5) and (3.1).   A similar procedure as that used to
obtain relations (3.10) and (3.11)
 can be employed to write down four of these remaining relations.
 We carry this out in the Appendix.
Like with (3.10) and (3.11), we find two distinct solutions,
and they correspond to case a) and case b).  For case a) we find:
\ba
 x\;\delta \dot y &=&q\;\delta \dot y \;x
+(pq-1)\;(\delta \dot x \; y +\delta x \; \dot y)\;,\no \\
 y\;\delta \dot x &=& p\;\delta \dot x \; y+(pq-1)\;\delta y\;\dot x \;,
 \no\\
 \dot x\;\delta y &=&q\;\delta y\;\dot x\;, \\
 \dot y\;\delta x &=&(1-pq )\;\delta y\;\dot x+p\;\delta x\;\dot y\;,\no
 \ea
and for case b) we find:
\ba
 x\;\delta \dot y &=&q\;\delta \dot y \;x
+(pq-1)\;(\delta \dot x \; y +{1\over p}\;\delta y\;\dot x)
+(pq+{1\over{pq}}-2)\;\delta x \; \dot y\;,\no\\
 y\;\delta \dot x &=& p\;\delta \dot x \; y+
 {1\over q}(pq-1)\;\delta x\;\dot y\;,\no\\
 \dot x\;\delta y &=&{1\over p}\;\delta y\;\dot x
+{1\over {pq}}(pq-1)\delta x\;\dot y\;,\\
 \dot y\;\delta x &=& {1\over q}\;\delta x \; \dot y\;.\no
 \ea
Both sets of relations (3.12) and (3.13)
satisfy the identities (3.2), and
are preserved under the $GL_{q,p}(2)$ transformations (1.3).
By taking the exterior derivative of eqs. (3.12) and (3.13),
we recover the relations (3.10) and (3.11), respectively.

The final four commutation relations are
between velocities $\dot x$ and $\dot y$
and their exterior derivatives
$\delta \dot x$ and $\delta \dot y$.  To obtain them
we first note that the commutation relation
between the two coordinates $x$ and $y$ of $Q_{q,p}$
is the same as that between the two velocities
$\dot x$ and $\dot y$, and the commutation relation
between the one forms $\delta x$ and $\delta y$ is
the same as that between
$\delta \dot x$  and $\delta \dot y$.  Then from (1.5),
a self-consistent set of
commutation relations between velocities and their exterior derivatives
is
\ba
\dot x\;\delta \dot x &=& pq\; \delta \dot x\;\dot  x \;,    \no\\
\dot x\;\delta \dot y &=& q\;\delta \dot y\;\dot  x
+ (pq-1)\; \delta \dot x\; \dot y \;,    \no\\
\dot y\;\delta \dot x&=& p\; \delta \dot  x\;\dot  y \;,\\
\dot y\;\delta \dot y&=& pq\; \delta \dot y\;\dot  y \;. \no
\ea
It can also be checked that these relations are consistent with
(3.12) and (3.13), in that no secondary conditions arise from commuting
$z_i$ through (3.14).

To summarize we have found two consistent differential calculi on
$TQ_{q,p}$:  case a) which is specified by eqs.
(1.2), (1.5), (3.1), (3.3), (3.5), (3.10), (3.12) and (3.14), and
case b) where (3.11) and (3.13) are substituted for (3.10) and (3.12).
In both cases the usual differential calculus on a tangent plane
is recovered when $q=p=1$.

\sxn{ Lagrangian and Hamiltonian Mechanics on
$Q_{q,p}$}

In order to define a Lagrangian formalism for particles
moving on $Q_{q,p}$
we need to take partial derivatives with respect to
the coordinates of the tangent quantum plane $TQ_{q,p}$.
Here we shall work exclusively in terms of right derivatives.
If $K$ is a function on the quantum plane $Q_{q,p}$,
we define the right derivatives of $K$ by writing
 variations $\delta K$ of $K$ according to
$\delta K(x,y) = \delta x {{\partial K}\over {\partial x}}+
\delta y {{\partial K}\over {\partial y}}  \;.$
The Lagrangian $L$ should be a function on the tangent plane $TQ_{q,p}$.
Right derivatives of $L$ are defined by
\be
\delta L(x,y,\dot x,\dot y) =
 \delta x {{\partial L}\over {\partial x}}+
\delta y {{\partial L}\over {\partial y}} +
 \delta \dot x {{\partial L}\over {\partial \dot x}}+
\delta\dot y {{\partial L}\over {\partial \dot y}}
 \quad.
\ee

An action principle can now be formulated on $TQ_{q,p}$,
and it leads to the
standard form for the Euler-Lagrange equations (provided we interpret
all derivatives with respect to the coordinates of $TQ_{q,p}$
as right derivatives).  If we define an action
$S=\int dt \; L(x,y,\dot x,\dot y) $, then we obtain
\be
 \dot\pi_x  - {{\partial L}\over {\partial x}}\;  \;  =\;  \;
 \dot\pi_y  - {{\partial L}\over {\partial y}}  \;\;  =\;0
\ee
when we ``extremize" the action, that is set $\delta S=0$.
$\pi_x$ and $\pi_y$ are the canonical momenta
associated with $x$ and $y$ respectively, where we assume the usual
definition,
\be
\pi_x = {{\partial L}\over {\partial\dot x}}
\quad {\rm and}\quad
\pi_y = {{\partial L}\over {\partial\dot y}} \;\;.
\ee

The solutions of the equations of motion (4.2) must be consistent
with the commutation relations (1.1) and (2.1-3), and as we will
see in example 1, this may not be possible.

In order to pass to the Hamiltonian formalism we need
 to define a phase space $T^*Q_{q,p}$.  It should be
  spanned by the variables $x$, $y$, $\pi_x$ and $\pi_y$.
We thus need to know the commutativity properties of
the coordinates and momenta.  But the commutativity
properties of the momenta can only be determined once we know
how to write $\pi_x$ and $\pi_y$ in terms of
$x$, $y$, $\dot x$ and $\dot y$ from eqs. (4.3),
For this we must know the functional form of $L$.
So the commutativity properties of the phase space variables are
dynamically determined from the initial choice of Lagrangian.

To write down Hamilton's equations of motion for the system,
we further need the differential calculus on
 $T^*Q_{q,p}$, and this too can only be determined
after knowing $\pi_x$ and $\pi_y$ in terms of
$x$, $y$, $\dot x$ and $\dot y$.  To be more explicit, let us define the
Hamiltonain according to
\be
H= \dot x\;\pi_x  + \dot{y}\;\pi_y -L \;\;,
\ee
and note from (4.1) and (4.2)
that variations $\delta H$ of $H$ can be written
\be
\delta H = -\delta x {{\partial L}\over {\partial x}}
-\delta y {{\partial L}\over {\partial y}}
+\dot x\;\delta\pi_x  + \dot{y}\;\delta\pi_y   \quad .
\ee
In terms of right derivatives of $H$, we
can also express the variations $\delta H$ according to
\be
\delta H =  \delta x {{\partial H}\over {\partial x}}  +
\delta y {{\partial H}\over {\partial y}}  +
\delta \pi_x {{\partial H}\over {\partial \pi_x}}+
\delta \pi_y {{\partial H}\over {\partial \pi_y}}\;.
\ee
Hamilton's equations of motion are standardly obtained by
equating the right hand sides of eqs. (4.5) and (4.6),
and by assuming independent variations $\delta x$, $\delta y$,
 $\delta \pi_x$ and $\delta \pi_y$.   But for this we need to know
the commutativity properties of $\dot x$ with $\delta\pi_x$ and
$\dot y$ with $\delta\pi_y$.  Thus the form of Hamilton's
equations of motion is dependent on the initial choice of
Lagrangian.  Moreover, the results
may be different for the case a) and the case b) commutation relations.
In general, we cannot even conclude from (4.5) and (4.6) that
$\dot\pi_x =- {{\partial H}\over {\partial x}} $ or
$\dot\pi_y =- {{\partial H}\over {\partial y}} $.

Also dependent on the choice of $L$ is the form of the
``deformed Poisson bracket" $\{\;,\;\}_{q,p}$.
This must be true if one requires that the time evolution of
any function ${\cal F}$ of
 $x$, $y$, $\pi_x$ and $\pi_y$ (and possibly $t$) should
be determined from the equation
\be
 \dot{\cal F}=
\{{\cal F}, H\}_{q,p} + {{\partial {\cal F}}\over{\partial t}} \;.
\ee
For us, eq. (4.7) defines the deformed Poisson bracket.
For arbitrary values of $q$ and $p$,
it will in general differ from the usual definition of the
Poisson bracket.

We next illustrate these ideas with two examples of Lagrangian systems
on $TQ_{q,p}$.

\centerline{{\bf Example 1}}

The first Lagrangian we consider
is second order in time derivatives and corresponds to a
deformation of the nonrelativistic particle (with mass
equal to one) in a scalar potential $V= V(x,y)$.  It is:
\be
L={1\over{1+pq}}(\dot x^2 + \dot y^2)- V(x,y)\;\;.
\ee
Using (3.14), variations $\delta L$ of $L$ are
\ba
\delta L&=&{1\over{1+pq}}(\delta \dot x\;\dot x + \dot x\;\delta \dot x
+\delta \dot y\;\dot y + \dot y\;\delta \dot y  )
-\delta x {{\partial V}\over {\partial x}}-
\delta y {{\partial V}\over {\partial y}}   \no\\
&=&\delta \dot x\;\dot x +\delta \dot y\;\dot y
-\delta x {{\partial V}\over {\partial x}}-
\delta y {{\partial V}\over {\partial y}}   \;
\ea
which leads to the usual  form for the Euler-Lagrange equations for a non
relativistic particle in a scalar potential
\be
\ddot x= -{{\partial V}\over {\partial x}}
\quad{\rm and}\quad\ddot y
= -{{\partial V}\over {\partial y}}\;\;.
\ee

For arbitrary values of $q$ and $p$,
the commutation relations (2.3) put strong constraints on the
allowable solutions to the equations of motion.  In Section 2, we
saw that they led to only trivial solutions to the equations for
a free particle.  Similar conclusions can be drawn for non zero
potentials $V$.  Eqs. (2.3) lead to a condition on $V$ itself:
\be
q\;{{\partial V}\over {\partial y}}\;x -
\; x\;{{\partial V}\over {\partial y}} =
\;{{\partial V}\over {\partial x}}\;y -
q\; y\;{{\partial V}\over {\partial x}}  \;\;.
\ee
When $q=p=1$, this is satisfied for any commuting function $V$ of
$x$ and $y$.

A $V$ which satisfies (4.11)
for arbitrary values of $q$ and $p$, is the harmonic oscillator potential
\be
V(x,y)= {{ x^2 + y^2}  \over{1+pq}}  \;\;.
\ee
The only other potentials satisfying (4.11) for $pq\ne 1$ are those
obtained by multiplying
both terms in (4.12) by c-number coefficients.  (More possibilities arise
when one introduces noncommuting constants in the theory, as is done
in ref.\cite{are}.)   Even these systems have no nontrivial solutions
because the remaining conditions in (2.3) impose additional
 constraints between the velocities and coordinates.  For the potential
(4.12) we get:  $\dot x^2=x^2$, $\dot y^2=y^2$ and $\dot x \dot y= xy$.

For the case of $qp=1$, nontrivial solutions to the equations of motion
(4.10) are possible.  In that case, all of equations (2.3) can be written
\ba
x\;{{\partial V}\over {\partial x}}=
{{\partial V}\over {\partial x}}\;x
 \;,  &\qquad &
x\;{{\partial V}\over {\partial y}}= q\;
{{\partial V}\over {\partial y}}\;x
 \;,  \no\\
y\;{{\partial V}\over {\partial y}}=
{{\partial V}\over {\partial y}}\;y
\;, &\qquad &
q\;y\;{{\partial V}\over {\partial x}}=
{{\partial V}\over {\partial x}}\;y
\; \;,
\ea
and there are no further constraints on the velocities and coordinates.
Conditions (4.13) are, for instance, satisfied for the potential
(4.12).  When $pq=1$, the commutation relations of the previous
sections simplify significantly and cases a) and b) coincide.
Below we shall see shall not restrict $pq$ to be $1$.
 For example 1, we will see that Hamilton's equations of motion
and the Poisson brackets have the usual form when $qp=1$.

{}From eq. (4.3), the canonical momenta $\pi_x$ and $\pi_y$
are identified with the velocity components
\be
\pi_x=\dot x \quad{\rm and}\quad \pi_y=\dot y\;.
\ee
This then defines the commutativity properties of the phase space
variables, and also defines the differential calculus on
$T^*Q_{q,p}$, this is since
the commutation relations for $\pi_x$ and $\pi_y$ (and their variations)
must be identical to the commutation relations for
$\dot x$ and $\dot y$ (and their variations), respectively.

{}From (4.4), the Hamiltonian for the system is
\be
H={{pq}\over{1+pq}}(\pi_x^2 + \pi_y^2)  +  V(x,y)\;\;.
\ee
Variations $\delta H$ of $H$ can now be written
\be
\delta H=  -\delta x {{\partial L}\over {\partial x}}
-\delta y {{\partial L}\over {\partial y}}
+pq\;\delta \pi_ x \;\dot x  + pq \;\delta\pi_y\;\dot y   \;,
\ee
where we have used (3.14) and (4.5).
By comparing eq. (4.16) with (4.6), we then get the following Hamilton's
equations of motion
\ba
\dot\pi_x =- {{\partial H}\over {\partial x}} \;,  &\qquad &
\dot x ={1\over{pq}}\; {{\partial H}\over {\partial \pi_x}} \;,  \no\\
\dot\pi_y =- {{\partial H}\over {\partial y}}\;, &\qquad &
\dot y ={1\over{pq}} \;{{\partial H}\over {\partial \pi_y}}\; \;.
\ea
Using $H$ given in eq. (4.15) we can verify that Hamilton's
equations of motion (4.17) for the system are identical
to the Euler-Lagrange equations (4.10).  Eqs. (4.17) reduce
to the usual form for Hamilton's equations of motion when $qp=1$.

If ${\cal F}$ is an arbitrary function of $x$, $y$, $\pi_x$, $\pi_y$
 and $t$, its time derivative can then be written according to:
\ba
 \dot{\cal F}&=&
\dot x {{\partial {\cal F}}\over {\partial x}}  +
\dot y {{\partial {\cal F}}\over {\partial y}}  +
\dot \pi_x {{\partial {\cal F}}\over {\partial \pi_x}}+
\dot \pi_y {{\partial {\cal F}}\over {\partial \pi_y}}+
 {{\partial {\cal F}}\over {\partial t}}  \no\\
 &=&{1\over{pq}}\; {{\partial H}\over {\partial \pi_x}}
\; {{\partial {\cal F}}\over {\partial x}}
+{1\over{pq}} \;{{\partial H}\over {\partial \pi_y}}
\; {{\partial {\cal F}}\over {\partial y}}
- {{\partial H}\over {\partial x}}
\;{{\partial {\cal F}}\over {\partial \pi_x}}
- {{\partial H}\over {\partial y}}
\; {{\partial {\cal F}}\over {\partial \pi_y}}
+ {{\partial {\cal F}}\over {\partial t}} \;\;,
\ea
where we have applied Hamilton's equations of motion.
$\dot{\cal F}$ can be written in the form (4.7) if we define
the deformed Poisson bracket $\{\;,\;\}_{q,p}$ of two functions
${\cal F}$ and ${\cal G}$ of the phase space variables
 $x$, $y$, $\pi_x$ and $\pi_y$ as follows:
\be
\{{\cal F},{\cal G}\}_{q,p}=
 {1\over{pq}}\; {{\partial {\cal G}}\over {\partial \pi_x}}
\; {{\partial {\cal F}}\over {\partial x}}
- {{\partial {\cal G}}\over {\partial x}}
\;{{\partial {\cal F}}\over {\partial \pi_x}}   \;\;
+\;\; {1\over{pq}}\; {{\partial {\cal G}}\over {\partial \pi_y}}
\; {{\partial {\cal F}}\over {\partial y}}
- {{\partial {\cal G}}\over {\partial y}}
\;{{\partial {\cal F}}\over {\partial \pi_y}}   \;\;  .
\ee
As a result of this definition, the Poisson brackets
of the phase space variables will not be antisymmetric for $pq \ne 1$.
  The canonical Poisson bracket relations are
$$
\{x,\pi_x\}_{q,p}=\{y,\pi_y\}_{q,p}= {1\over{pq}}  \;,
$$
\be
\{\pi_x, x\}_{q,p}=\{\pi_y,y\}_{q,p}= -1   \;.
\ee
If we identify these Poisson brackets with elements of a matrix
whose inverse corresponds to the symplectic two form
$\omega$, the symplectic two form then takes the form
$\omega_{q,p} = (pq+1)\;(\delta \pi_x\wedge\delta x
+\delta \pi_y\wedge\delta y )   \;,   $
where we have used the commutation relations (3.3).
We see that the usual symplectic structure is recovered for $pq=1$.

To compute the deformed Poisson brackets (4.19) of functions of
the phase space variables (which we can write in terms of a formal power
series), we need to take partial derivatives
of products of functions ${\cal G}$
and $  {\cal H}$ on $T^*Q_{q,p}$.  For this purpose
it is necessary to know the commutativity properties
of such functions with the phase space variables (or more precisely,
with variations of the phase space variables).
If we denote the phase space variables by $\tilde z_i$, then
we can specify these commutativity properties by
\be
{\cal G}\;\delta \tilde z_i =\delta \tilde z_j \;{\cal O}^{({\cal G})}
_{ji}\;.
\ee
${\cal O}^{({\cal G})}_{ij} $ can be computed using (4.14) and
the commutation relations of Sections 1 and 3.
Although the usual Leibniz rule is
assumed to hold for the exterior derivative $\delta$,
it does not necessarily
apply for derivatives with respect to the phase space
variables.  Rather, from $\delta ({\cal GH})=
\delta {\cal G}\;{\cal H}+  {\cal G}\;\delta {\cal H}$ and eq.
(4.21), one gets\cite{wz}
\be
 {{\partial ({\cal GH}})\over {\partial \tilde z_i}}
={{\partial {\cal G}}\over {\partial \tilde z_i}}  \;{\cal H} \;+\;
 {\cal O}^{({\cal G})}_{ij} \;
{{\partial {\cal H}}\over {\partial \tilde z_j}} \;\;.
\ee

\centerline{{\bf Example 2}}

The second Lagrangian we consider
is first order in time derivatives and its bosonic and fermionic
analogues have been studied long ago\cite{cas}.   It is
\be
L= {{x\;\dot y-q\;y\;\dot x}\over{1+pq}}\;.
\ee
(A similar Lagrangian was examined in ref. \cite{mm} but there
a different differential calculus was used.)
Up to a total time derivative, $L$ is equivalent
to  $  x\;\dot y \;.$   The Lagrangian (4.23)
 has the property that when we restrict
to the case $q=p$, it is invariant under a subset
of the $GL_{q,p}(2)$ transformations defined in eqs. (1.3) and (1.4).
This subset is the one parameter deformation of the special linear
group in two dimensions, standardly
denoted by $SL_{q}(2)$, and it is obtained by setting
\be
q=p\quad{\rm and  \quad det}_q \;  [ T ]  =1\;\;,
\ee
where
$$
{ \rm det}_q \;  [ T ] \equiv AD-qBC\;\;.
$$
Here one notes that ${\rm  det}_q [ T ]$
 so defined commutes with all matrix elements $A$, $B$,
$C$ and $D$ when $q=p$, and therefore can be identified with the
c-number ``one".

By taking the variational derivative of $L$ we can get two different
answers, depending upon whether we use case a) or case b)
commutation relations:
\ba
(1+pq)\;\delta L&=&\delta x\;\dot y+
 x\;\delta\dot y-q\;\delta y\;\dot x -q\;y\;\delta\dot x  \no\\
&=&\left\{
\matrix{pq\;\delta x \; \dot y + q\;\delta\dot y\; x-\delta\dot x\; y
-pq^2\;\delta y\;\dot x\;, & {\rm case \; a)},\cr  &\cr
{1\over{pq}}\;\delta x \; \dot y + q\;\delta\dot y\; x-\delta\dot x\; y
-{1\over p}\;\delta y\;\dot x\;, & {\rm case \; b)}.\cr} \right.
 \ea
In either case however we obtain the trivial equations of motion
$  \dot x =\dot y = 0 \; $,
 and as a result the Lagrangian vanishes ``on mass shell".

The equations for the canonical momenta correspond to primary
constraints (in the sense of Dirac), and they have the same
form for both case a) and b),
\be
\phi_x\equiv\pi_x+{y\over {1+pq}}\approx 0 \quad {\rm and}\quad
\phi_y\equiv\pi_y-{{q\; x}\over {1+pq}}\approx 0   \quad.
\ee
As a result, the commutativity properties of $\pi_x$ and $\pi_y$
(and their variations) are identical to the commutativity properties of
$-{y\over{1+pq}}$ and ${{q\;x}\over{1+pq}}$
(and their variations), respectively.

It is easy to check that the Hamiltonian for this system is zero, or
more precisely, it is a linear combination of the constraints (4.26).
Now variations $\delta H$ of $H$ are different for the cases
a) and b).  From eq. (4.5),
\be
\delta H=  -\delta x {{\partial L}\over {\partial x}}
-\delta y {{\partial L}\over {\partial y}}
+{{s^2 q^2}\over p}\;\delta \pi_ x \;\dot x
+{{s^2}\over {p}} \;\delta\pi_y\;\dot y   \;,
\ee
where $s$ can take two values: $s=p$ and $s={1\over q}$.
The former corresponds to case a) and the latter corresponds to case b).
As a result of (4.27), Hamilton's equations of motion
will be different for the two cases:
\ba
\dot\pi_x =- {{\partial H}\over {\partial x}} \;, &\qquad &
\dot x ={p\over{s^2 q^2}}\; {{\partial H}\over {\partial \pi_x}} \;,\no\\
\dot\pi_y =- {{\partial H}\over {\partial y}} \;,
&\qquad &\dot y ={{p}\over{s^2}}\;{{\partial H}\over{\partial \pi_y}}\;.
\ea

If ${\cal F}$ is an arbitrary function of
 $x$, $y$, $\pi_x$, $\pi_y$ and $t$, then its time derivative
$\dot{\cal F}$ can be written in the form (4.7) if we define
the deformed Poisson bracket $\{\;,\;\}_{q,p}$ of two functions
${\cal F}$ and ${\cal G}$ on the phase space as follows:
\be
\{{\cal F},{\cal G}\}_{q,p}=
 {p\over{s^2 q^2}}\; {{\partial {\cal G}}\over {\partial \pi_x}}
\; {{\partial {\cal F}}\over {\partial x}}
- {{\partial {\cal G}}\over {\partial x}}
\;{{\partial {\cal F}}\over {\partial \pi_x}}   \;\;
+\;\; {{p}\over{s^2}}\; {{\partial {\cal G}}\over {\partial \pi_y}}
\; {{\partial {\cal F}}\over {\partial y}}
- {{\partial {\cal G}}\over {\partial y}}
\;{{\partial {\cal F}}\over {\partial \pi_y}}   \;.
\ee
Then the Poisson brackets of the phase space variables will be
$$
\{x,\pi_x\}_{q,p}=  {p\over{s^2 q^2}}  \;, \qquad
\{y,\pi_y\}_{q,p}=  {{p}\over{s^2}}  \;, $$
\be
\{\pi_x, x\}_{q,p}=\{\pi_y,y\}_{q,p}= -1   \;.
\ee
Applying these Poisson brackets, we see that the constraints are not
first class [neither for case a) or for case b)], since
\be
\{\phi_x,\phi_y\}_{q,p}=  {1\over{s}}  \;, \qquad
\{\phi_y,\phi_x\}_{q,p}= - {1\over{sq}} \; .
\ee
(Here we assume $pq\ne -1$.)

To eliminate the constraints $\phi_x$ and $\phi_y$, we
can apply the analogue of the Dirac procedure and replace deformed
Poisson brackets $\{\;,\;\}_{q,p}$ by deformed
Dirac brackets $\{\;,\;\}_{q,p}^*$.  We define the latter by
\be
\{{\cal F},{\cal G}\}_{q,p}^*=
\{{\cal F},{\cal G}\}_{q,p}+s\;\biggl( q\;
\{{\cal F},\phi_x \}_{q,p}\;\{\phi_y,{\cal G}\}_{q,p}-  \;
\{{\cal F},\phi_y \}_{q,p}\;\{\phi_x,{\cal G}\}_{q,p}\biggr) \;,
\ee
where ${\cal F}$ and ${\cal G}$ are arbitrary functions on
$T^*Q_{q,p}$.  With this definition, it follows that Dirac
brackets of the constraints $\phi_x$ and $\phi_y$ with any functions
on the phase space vanish.  We can now eliminate the constraints
from the theory by working on a reduced phase space.  We can take
the reduced phase space to be the original quantum plane $Q_{q,p}$
parametrized by $x$ and $y$.  If $F$ and $G$ are functions on
the reduced phase space, their Dirac brackets simplify to
\ba
\{ F, G\}_{q,p}^*&=&s \;
\biggl(q\;\{ F,\pi_x \}_{q,p}\;\{\pi_y, G\}_{q,p}-
\{ F,\pi_y \}_{q,p}\;\{\pi_x, G\}_{q,p}\biggr) \no\\
&= & {p\over{s}}\;\biggl( -{1\over q}\;
 {{\partial F}\over {\partial x}}
\; {{\partial G}\over {\partial y}}
+\;{{\partial F}\over {\partial y}}
\; {{\partial G}\over {\partial x}}  \biggr)\;\;.
\ea
We then see that the Dirac brackets between the coordinates $x$ and $y$
do not vanish, and further, are not antisymmetric.  Instead,
\be
\{x,y\}_{q,p}^*= - {p\over{sq}}  \;, \qquad
\{y,x\}_{q,p}^*=  {{p}\over{s}}  \;.
\ee

The Dirac bracket relations (4.34) are preserved under $SL_q(2)$
transformations [cf. eqs. (1.3) and (4.24) ].  Such transformations
are therefore $canonical$.

Instead of being antisymmetric, the brackets (4.34) have the property:
\be
\{x,y\}_{q,p}^*= - {1\over{q}}  \;\{y,x\}_{q,p}^*  \;.
\ee
Eq. (4.35) holds for both case a) and case b).  From (4.34)
the symplectic two form for the theory is proportional to the two form
on $Q_{q,p}$:
$$
\omega_{q,p} = {s\over p}(pq+1)\;\delta x\wedge\delta y \;.
$$

\sxn{Concluding Remarks}

In Section 3 and the Appendix we were able to construct
 two different differential calculi on $TQ_{q,p}$.
The significance of having two different calculi is not
evident, and there seems to be no reason for preferring one
over the other.  From them, it appears that starting from a single
Lagrangian one
 may, in principle, derive two distinct dynamical systems.
Furthermore, it may be possible to construct more differential calculi,
and hence more dynamical systems,
if we start with more general ans\"atze then the ones we used.

Although we developed the Lagrangian and Hamiltonian formalisms
for particles moving on $Q_{q,p}$,
 we were unable to find nontrivial dynamical systems for
arbitrary values of $p$ and $q$ which were consistent with the
commutation relations (2.3).  Further,
the Lagrangians we examined in Section 4 were not central elements
of the algebra generated by $x$, $y$, $\dot x$ and $\dot y$.
They thus cannot be identified with c-numbers.  (However on mass shell,
the Lagrangian (4.22) was a c-number, namely zero.)   The interpretation
of Feynman path integrals
 using such Lagrangians thus becomes problematic.

Nevertheless, a consistent canonical quantization of these
systems may be possible, even though the Lagrangians are not c-numbers.
 The quantization procedure however is not uniquely determined.
 For example 2, we want to replace $x$ and $y$, by some new noncommuting
operators ${\bf x}$ and ${\bf y}$, where the latter satisfy
(1.1) in the limit $\hbar \rightarrow 0$.  In the limit of $q=p=1$,
we should of course recover the usual quantization.
For example 2, a possible quantization could mean replacing
Dirac brackets (4.34) by the relation
\be
{1\over q}\;{\bf xy}- {\bf yx} =-{{i\hbar p}\over{sq}}   \;.
\ee
${\bf x}$ and ${\bf y}$ are then deformed creation and annihilation
operators.
{\it Thus for this particular example $x$ and $y$ are indeed the
classical limit of deformed creation and annihilation operators.}
But this result is not general as it was obtained specifically from
Lagrangian (4.23).

Finally, we remark that although the Lagrangian (4.23) is invariant
under $SL_q(2)$ transformations when $q=p$, and these transformations
 correspond to a canonical symmetry,
we do not know how to apply Noether's theorem in this case to
find the analogue of the symmetry generators (except of
course, for the case of $q=p=1$,
where the $SL(2)$ generators are $1$ and $i$ times binomials of $x$
and $y$).  The reason is that to apply Noether's theorem, we must be
able to write the symmetry transformation (1.3) in infinitesimal form,
and this is a non trivial problem when $q=p \ne 1$.  Yet $SL_q(2)$
(actually, $SU_q(2)$) generators have been constructed from a single
set of deformed creation and annihilation operators.\cite{bied2}

\axn{Appendix}

Here we derive the commutation relations of
$ x$ with $\delta \dot y $, $y$ with $\delta \dot x $,
$\dot x$ with $\delta y$ and $\dot y$ with $ \delta x $.
For this, we define a $4\times 4$ matrix
$[  F  ] $, whose elements are c-numbers, and make the ansatz:
\be
 \pmatrix{ x\;\delta \dot y \cr
 y\;\delta \dot x \cr
 \dot x\;\delta y \cr
 \dot y\;\delta x \cr }
= [  F  ]  \;
 \pmatrix{\delta \dot y \;x\cr
\delta \dot x \;y\cr
\delta y \; \dot x \cr
\delta x \; \dot y \cr }    \quad .
\ee
The matrix elements can be determined I) by demanding
that no secondary conditions on combinations of
$z_i$ and $\delta z_i$ result from
commuting $z_i$ and $\delta z_i$ through the commutation relations
(1.1), (2.1) and (2.2), II) from the identities (3.2) and III)
by demanding that the relations (A.1) are
preserved under the $GL_{q,p}(2)$ transformations (1.3).

I) By multiplying $\delta \dot x$ and $\delta \dot y$ on the right of
$x\;y-q\;y\;x=0$, we find
\be
F_{11}=q \quad {\rm and}\quad F_{21}=0\;,
\ee
along with the consistency conditions
\ba
F_{12}(p-F_{22})&=&0\;,\no\\
(1-pq)(pF_{13} +1- pq) &=& F_{12}F_{23}\;,    \no\\
p(1-pq)(F_{14} +1- pq) &= &F_{12}F_{24}\;,    \\
F_{22} +p(F_{23}-pq) +F_{24}pq &=&0\;.\no
\ea
By multiplying $\delta  x$ and $\delta  y$ on the right of
the commutation relations (2.1), we find that
\be
 F_{31} =0\quad{\rm and}\quad F_{41}=0\;,
\ee
along with the consistency conditions
\ba
 F_{42} + pF_{32}&=&0\;,\no\\
F_{42}F_{24}=F_{42}F_{23}&=&0\;,\no\\
F_{42}(F_{22}-p^2 q)&=&0\;,\no\\
1-F_{43} - pF_{33}&=&0\;,\\
p-F_{44} - pF_{34}&=&0 \;,\no\\
F_{32} + pF_{33} +pq(F_{34}-1)&=&0\;,\no\\
F_{42} + pF_{43} +pqF_{44}-p&=&0 \;.\no
\ea

II)  Upon substituting the ansatz (A.1) into the identities (3.2),
we get the following additional conditions:
\ba
F_{12}+F_{32}&=&qp-1\;,\no\\
F_{13}+F_{33}&=&q\;,\no\\
F_{14}+F_{34}&=&qp-1\;,\no\\
F_{22}+F_{42}&=&p\;,\\
F_{23}+F_{43}&=&0\;,  \no\\
F_{24}+F_{44}&=&p\;.\no
\ea

For arbitrary $q$ and $p$, there are two types of solutions to
equations (A.3), (A.5) and (A.6).
One of them yields the following matrix $ [  F  ] $:
\be
[  F  ]   = \pmatrix{
q & 0 & {1\over p}(pq-1) & pq-1 \cr
0 & p^2q & 0 & 0 \cr
0 & pq-1 & {1\over p} & 0 \cr
0 & p(1-pq) & 0 & p \cr} \;,
\ee
However, it is not hard to show that the commutation relations
resulting from (A.7) do not fulfill III), that is,
they are not preserved under $GL_{q,p}(2)$ transformations (1.3),
and we shall thus not consider this solution further.

On the other hand, $GL_{q,p}(2)$ transformations do preserve
the other type of solution:
\be
[  F  ]   = \pmatrix{
q & qp-1 & q(1-{s\over p}) & pq-2+{s\over p} \cr
0 & p & qs-1 & p-s \cr
0 & 0 & {{qs}\over p} & 1-{s\over p} \cr
0 & 0 & 1-qs & s \cr} \;.
\ee
To determine the quantity $s$, we can I) multiply $\delta  x$
 on the right of $\dot x\;\dot y-q\;\dot y\;\dot x=0$.  This yields:
\be
(1-qs)({s\over p}-1)=0\;.
\ee
$s$ is thus allowed to take two different values,
$s=p$ and $s={1\over q}$, and they correspond
to the two cases a) and b), respectively.
The commutation relations given in eqs. (3.12) and (3.13) result.


\newpage

\begin{thebibliography}{99}

\bibitem{rev} For reviews see,
L. Takhtajan in {\it Introduction to Quantum Group and Integrable Massive
Models of Quantum Field Theory}, M-L. Ge and B-H. Zhao (eds.) (World
Scientific, 1990);  S. Majid, Int. J. Mod. Phys. {\bf 5}
(1990) 1;  T. Tjin, Int. J. Mod. Phys. {\bf 7} (1992) 6175.

\bibitem{YG} C.N. Yang and M. Ge, {\it Braid groups and statistical
physics} (World Scientific, Singapore, 1989).

\bibitem{ow} D. Fivel, Phys. Rev. Lett. {\bf 65} (1990) 3361;
O. W. Greenberg, Phys. Rev. {\bf D43} (1991) 4111;
A. Lerda and S. Sciuto, ``Anyons and Quantum Groups", Torino
preprint DFTT 73/92.

\bibitem{man} Yu. I. Manin, Commun. Math. Phys. {\bf 123}
(1989) 169.

\bibitem{bf} L. Baulieu and E. G. Floratos, Phys. Lett.
{\bf 258B} (1991) 171.

\bibitem{bied} L. C. Biedenharn, J. Phys. {\bf A22} (1989) L873;
A. J. Macfarlane, J. Phys. {\bf A22} (1989) 4581;
C-P Sun and H-C Fu, J. Phys. {\bf A22} (1989) L983;

\bibitem{bied2}
M. Nomura and L. C. Biedenharn, J. Math. Phys. {\bf 33}  (1992) 3636.

\bibitem{cas} R. Casalbuoni, Il Nuovo. Cim. {\bf 33A}
(1976) 115; 389; A. P. Balachandran, P. Salomonson, B. Skagerstam
J. Winnberg, Phys. Rev {\bf D15} (1977) 2308.

\bibitem{soni} A. Schirrmacher, J. Wess and B. Zumino, Z. Phys.
 {\bf C49} (1991) 317; A. Schirrmacher, ``Multiparameter R-matrices
and their Quantum Groups", Max Planck preprint MPI-ph/91-24;
 S. K. Soni, J. Phys. {\bf A24} (1991)
 L169; L459.

\bibitem{wz} J. Wess and B. Zumino, Nucl. Phys. (Proc. Suppl.)
{\bf 18B} (1990) 302.

\bibitem{pb} D. G. Caldi, ``Q-Deformations of the Heisenberg
equations of motion", Fermi Lab preprint FERMILAB-CONF-91/210-T;
I. Ya. Aref'eva and I. V. Volovich, Phys. Lett. {\bf B268} (1991)
179;  S. V. Shabanov, J. Phys. {\bf A25} (1992) L1245;
``Quantum and classical mechanics of
 q-deformed systems", Bern preprint BUTP-92/24.

\bibitem{are} I. Ya. Aref'eva, ``Quantum Group Gauge Fields",
CERN preprint CERN-TH.6207/91.

\bibitem{mm} J. L. Matheus-Valle and M. A. R. Monteiro,
 Mod. Phys. Lett. {\bf A7} (1992) 3029.

\end{thebibliography}
\end{document}